\documentclass[fleqn,10pt]{wlscirep}

\usepackage{graphicx}
\usepackage{subcaption}

\title{Gate-tunable electron interaction \\ in high-$\kappa$
dielectric films}

\author[1+]{Svitlana\,Kondovych}
\author[1+]{Igor\,Luk'yanchuk}
\author[2,3,4+]{Tatyana\,I.\,Baturina}
\author[5*+]{Valerii\,M.\,Vinokur}
\affil[1]{University of Picardie, Laboratory of Condensed Matter Physics,
Amiens, 80000, France}
\affil[2]{University of Regensburg, Universit\"atsstra\ss{}e 31, Regensburg 93053, Germany}
\affil[3]{A.\,V.\,Rzhanov Institute of Semiconductor Physics SB RAS, 13 Lavrentjev Avenue, Novosibirsk 630090, Russia}
\affil[4]{Novosibirsk State University, Pirogova str. 2, Novosibirsk 630090, Russia}
\affil[5]{Materials Science Division, Argonne National Laboratory,  
9700 S. Cass Avenue, Argonne, Illinois 60637, USA}

\affil[*]{vinokour@anl.gov}

\affil[+]{these authors contributed equally to this work}

\begin{abstract}
The two-dimensional (2D) logarithmic character of Coulomb interaction between charges and the resulting logarithmic confinement is a remarkable inherent property of high dielectric constant (high-$\kappa$) thin films with far reaching implications. Most and foremost, this is
the charge Berezinskii-Kosterlitz-Thouless transition with the notable manifestation, low-temperature superinsulating topological phase. 
Here we show that the range of the confinement can be tuned by the
external gate electrode and unravel a variety of electrostatic interactions in high-$\kappa$ films.
Our findings open a unique laboratory for the in-depth study of topological phase transitions and a plethora of 
related phenomena, ranging from criticality of quantum metal- and superconductor-insulator transitions to the effects of charge-trapping and Coulomb scalability in memory nanodevices.
\end{abstract}
\begin{document}

\flushbottom
\maketitle

\thispagestyle{empty}

\section*{Introduction}

High dielectric constant or high-$\kappa$ 2D systems
enjoy an intense experimental and theoretical attention, see Ref.\,[\citeonline{osada2012two}] and references therein. 
The interest is motivated by high
technological promise of these systems for fabrication of nanoscale capacitor
components  and for design of the novel memory elements
and switching devices of enhanced performance. The high-$\kappa$ devices comprise unprecedentedly wide spectrum of physical systems 
ranging from traditional dielectrics and ferroelectrics to strongly
disordered thin metallic and superconducting films experiencing metal-insulator and superconductor-insulator
transitions, respectively\,\cite{BatVin2013,PhysRevLett.34.1627,PhysRevLett.46.375,PhysRevB.25.5578,yakimov1997metal,PhysRevB.62.R2255}. The profound application of the high-$\kappa$ sheets is the
charge trapping  elements for flash memory \cite{zhao2014} enabling the storage of the multiple bits in a single memory cell, thus overcoming the scalability limit
of a standard flash memory. The challenging task crucial to applications is establishing the effective tunability of charge-trapping memory (CTM) units allowing for controlling the strength and spatial scale of charge distribution. 

The major feature of high-$\kappa$ systems leading to their unique properties, is that the electric field induced by the trapped charge remains confined within the film. This ensures the electrostatic integrity and stability with respect to external perturbations and gives rise to the 2D character of the Coulomb interactions between the charges\cite{Rytova1967,chaplik1971charged,keldysh1979coulomb}. Namely, the potential produced by the charge, located inside the high-$\kappa$ sheet of thickness $d$, sandwiched between media with $\kappa _{a}$ and $\kappa _{b}$ permeabilities, exhibits the logarithmic distance dependence, 
$\varphi (\rho )\propto \ln
(\rho /\Lambda)$, extending till the fundamental screening length of the potential dimensional crossover, $\Lambda ={\kappa } d/(\kappa_a+\kappa_b)$.
A striking example of the 2D Coulomb behaviour is 
the phenomenon of superinsulation in strongly disordered superconducting films\,\cite{vinokur2008superinsulator,BatVin2013,baturina2008hyperactivated,kalok2012non}. 
There, in the
critical vicinity of the superconductor-insulator transition, the superconducting film acquires an anomalously high $\kappa$, the Cooper pairs interact according to the logarithmic law, and the system experiences the charge
Berezinskii-Kosterlitz-Thouless (BKT) transition into a
state with the infinite resistance. 
Another general consequence of the logarithmic Coulomb interaction, is that the high-$\kappa$ sheets exhibit 
the so-called phenomenon of the global Coulomb blockade resulting in a logarithmic scaling of characteristic energies of the system with the relevant screening length, which is the smallest of either $\Lambda$ or the lateral system size. In the Cooper pair insulator, this manifests as the logarithmic scaling of the energy controlling the in-plane tunneling
conductivity \cite{fistul2008collective,vinokur2008superinsulator,baturina2011nanopattern}. In the CTM element, this is the logarithmic scaling of its capacitance. 

The screening length is a major parameter controlling the electric properties of the high-$\kappa$ films. Thus, their  applications require reliable and simple ways of tuning $\Lambda$ which, at the same time, maintain robustness of the underlying dielectric properties of the system. As we show below, this is achieved by the clever location of the control gate. 
Adjusting the distance between the high-$\kappa$ film and the gate, we vary the screening length of the logarithmic interaction and obtain a wealth of the electrostatic behaviors at different spatial scales, enabling  to control the scalability and capacitance of the system. In what follows we describe the electrostatic properties of the generic high-$\kappa$ device with the tunable distance to the control gate. 

\section*{Model}

\begin{figure}[ht!]
\begin{center}
\includegraphics[width=0.9\textwidth]{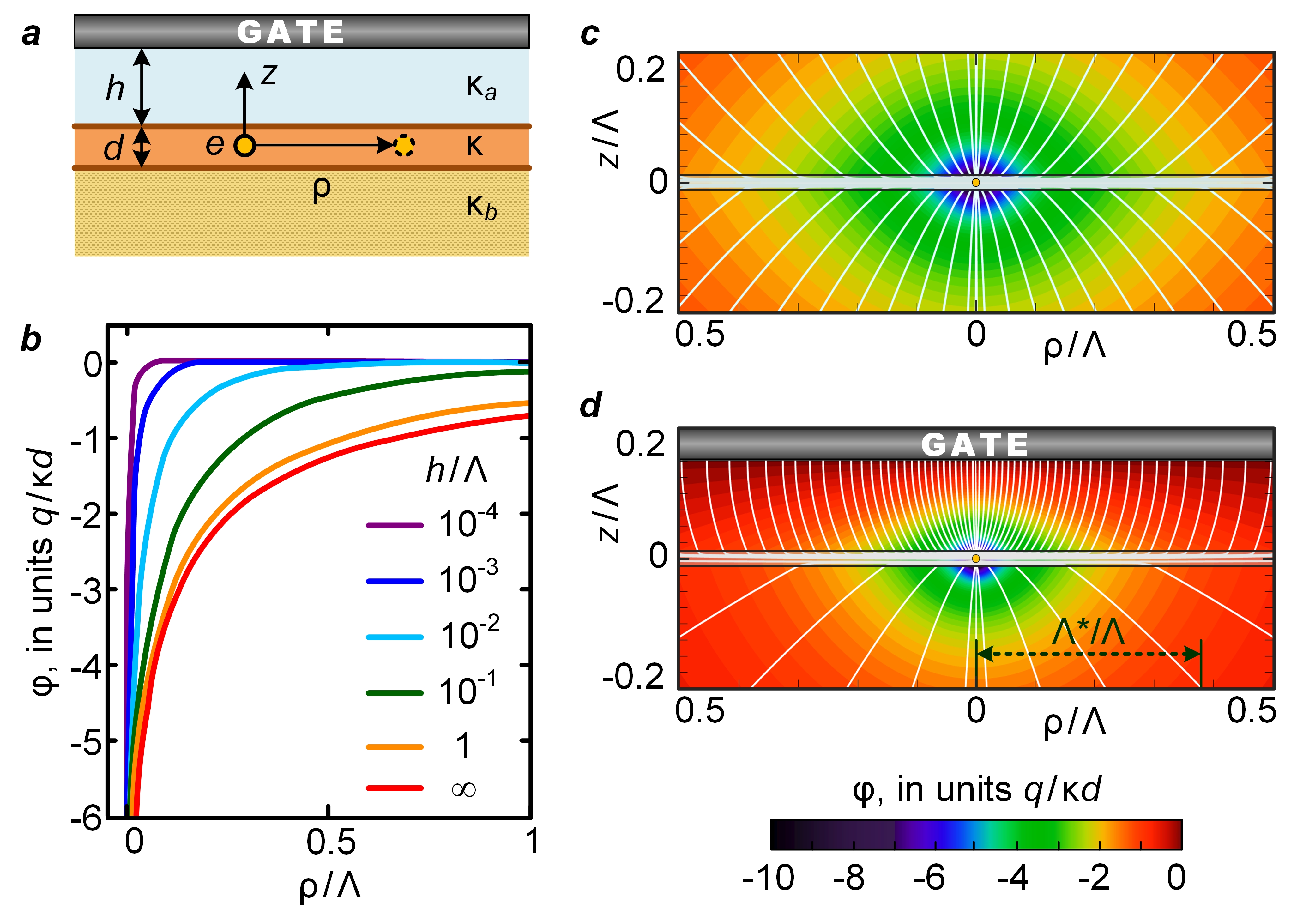}
\end{center}
\caption{\textbf{System geometry and spatial distribution of electrostatic
potential.} (a) Thin film of thickness $d$ with the dielectric constant $%
\protect\kappa$ is deposited on the substrate with the dielectric constant $%
\protect\kappa_b$. The metallic gate on top is separated from the film by
the spacer of thickness $h$ with the dielectric constant $\protect\kappa_a$.
Interacting charges, $e$, are located in the middle of the film. The origin
of the cylindrical coordinate system, $\rho,\theta,z$, with $%
\protect\rho$ being the lateral coordinate, is chosen at the location of the
charge generating the electric field; the $z$-axis is perpendicular to the
film plane. (b) The electrostatic potential, $\protect\varphi$, induced by
the charge $e<0$ as function of $\protect\rho$
for different distances $h$ between film and electrode. The values of $%
\protect\rho$ and $h$ are taken in units of the characteristic length $%
\Lambda$, the potential $\protect\varphi$ is taken in units $q/\kappa d$ where $%
q=e/4\protect\pi \protect\varepsilon _{0} $ and $\protect%
\varepsilon _{0}$ is the vacuum permittivity. The curves are calculated for $%
\protect\kappa=10^4$, $\protect\kappa_a=1$, $\protect\kappa_b=4$. (c) and
(d) Electric field lines (white) and the color map of the
electrostatic potential induced by charge $e<0$ in the
cross-sectional plane. Panel (c) displays the field and potential
without the gate; panel (d) shows the same in the presence of the gate. In
the panels (c) and (d) we take $\protect%
\kappa=100$, $\protect\kappa_a=1$, $\protect\kappa_b=1$.}
\label{fig:model}
\end{figure}

We consider a point charge, $e<0$, located in the middle of a high-$\kappa$
film of the thickness $d$, deposited on a dielectric substrate with the
dielectric constant, $\kappa _{b}$. A metallic gate is separated from the
film by a layer of the thickness $h$ with the dielectric constant $\kappa_a$, see Fig.\ref{fig:model}a. 

The origin of the cylindrical coordinate system with the $z$-axis
perpendicular to the film's plane, $\left( \rho ,\theta ,z\right) $, is
placed at the charge location (Fig.\ref{fig:model}a). 
In very thin films, 
which are the main focus of our study, we disregard the distances smaller than the film thickness and thus consider $\rho>d$. The relevant physical characteristic scale controlling the electrostatic properties of the system is the screening length $\Lambda$.  Then the Poisson
equations defining the potential distribution created by the charge assumes
the form: 
\begin{eqnarray}
\frac{1}{\rho }\partial _{\rho }\left( \rho \partial _{\rho }\varphi \right)
+\partial _{z}^{2}\varphi =4\pi \frac{ q}{\kappa}\delta_3(\rho,z), &&\quad
\left\vert z\right\vert <d/2,  \label{Poisson} \\
\frac{1}{\rho }\partial _{\rho }\left( \rho \partial _{\rho }\varphi
_{a,b}\right) +\partial _{z}^{2}\varphi _{a,b}=0, &&\quad \left\vert
z\right\vert >d/2\,.  \notag
\end{eqnarray}%
Here $\varphi $ is the electric potential inside the film, $\varphi _{a}$
and $\varphi _{b}$ are the potentials in the regions above and below the
film, respectively,  $\delta_3(\rho,z)= \delta(\rho)\delta(z)/2\pi\rho$
is the 3D Dirac delta-function in the cylindrical coordinates, $q=e$ and $q=e/4\pi \varepsilon _{0}$ in CGS and SI systems
respectively, $\varepsilon _{0}$ is the vacuum permittivity. The
electrostatic boundary conditions are $\varphi =\varphi _{a,b}$ and $\kappa
\partial _{z}\varphi =\kappa _{a,b}\partial _{z}\varphi _{a,b}$ at $z=\pm
d/2 $ at the film surfaces, and $\varphi _{a}=0$ at $z=h+d/2$ at the interface with the electrode. Then, the energy of the
interaction with the second identical electron located at the distance $\rho 
$ (see Fig.\ref{fig:model}a) is given by $U\left( \rho \right) =2e\varphi
\left( \rho \right) $. For numerical calculations we use typical values of
parameters for a InO film deposited on the SiO$_{2}$ substrate: the film
dielectric constant, $\kappa \simeq 10^{4}$, the substrate dielectric
constant, $\kappa _{b}=4$, and the dielectric constant for the air gap
between the film and the gate, $\kappa _{a}=1$, see Ref.\,[%
\citeonline{BatVin2013}].


\begin{figure}[b!]
\begin{center}
\includegraphics[width=0.9\textwidth]{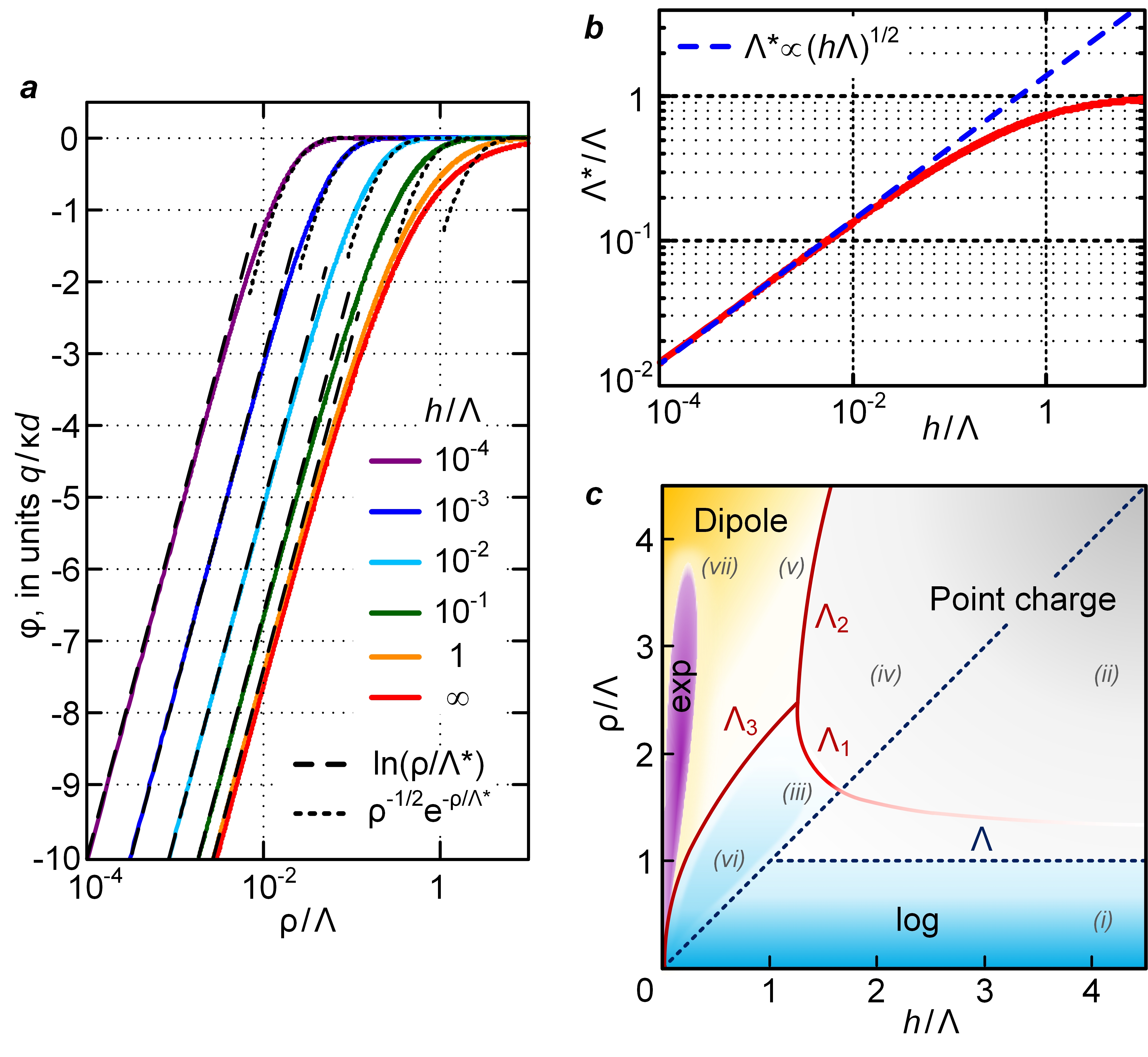}
\end{center}
\caption{\textbf{The electrostatic potential in the presence of the gate and the sketch of the regimes of electrostatic interactions.} The material dielectric parameters are the same as in Fig.\,\ref{fig:model}b. The distances are measured in units of the fundamental screening length $\Lambda$ and the potential in units $q/\kappa d$. (a) Semi-log plots of the electrostatic potential of the point charge placed in the middle of the film as functions of the distance for various values of the spacer, $h/\Lambda$, increasing from the top to the bottom. The straight dotted lines are fits to $\propto \ln(\rho/\Lambda^{\ast})$ dependencies at small distances from which we determine the screening lengths $\Lambda^\ast$ at different $h$.  
The dashed lines stand for the 
$\propto\rho^{-1/2}\exp(-\rho/\Lambda^\ast)$ dependencies, which provide pretty fair fits for the long-distance behaviour of $\varphi(\rho )$ at small $h\lesssim 10^{-2}\Lambda$. 
(b) The log-log plot of the $\Lambda^{\ast}$ on $h$ dependence determined from the data given in panel (a). 
At small separations between the gate and the film, $h\lesssim 10^{-2}\Lambda$, the effective screening length follows the law $\Lambda^{\ast}\simeq\sqrt{\Lambda h}$, at larger $h$ the noticeable deviation from this dependence is observed and at $h\gtrsim \Lambda$ it tends to $\Lambda$.
(c) The map visualizing the different interaction regimes between charges in the $h-\rho$ coordinates.  
The gate-dominated regime takes place at $\rho<h$, i.e. above the dashed diagonal line. 
Below this line the interaction is only slightly affected by the gate. The regions with the logarithmic interaction, lying at small $\rho$ are highlighted by the blueish colours. This 2D logarithmic interaction becomes screened at distances beyond the screening length. The latter can acquire either of the values $\Lambda$, $\Lambda_1$ or $\Lambda_3$, depending on the parameters of the system.  
In the screened regime, the charges interact either as 3D point charges (grayish region, on the right of the separating line $\Lambda_2$) or as the gate-imaged electric dipoles (yellowish region, on the left of $\Lambda_2$).  At very small gate separation the strong exponential screening  takes place (the violet petal).  Gray roman numerals indicate the correspondence to  analytical formulae in Table~I.   
}
\label{fig:potential}
\end{figure}


\section*{Results}

Results of the numerical solution to Eqs.\,(\ref{Poisson}) are shown in
Fig.\thinspace\ref{fig:model}b-d. The space coordinates are measured in
units $\Lambda $ defined in the Introduction. Panels (c) and (d) illustrate
the cross-section of the configuration of the electric field lines and the color 
map of the electrostatic potential for two characteristic cases, without and with
metallic gate respectively. For illustration purposes we assumed in panels (c) and (d) $%
\kappa =100$ and symmetric properties of the upper and lower dielectric
media, $\kappa _{a}=\kappa _{b}$. It can be immediately seen that
introducing the gate localizes potential within the smaller $h$-dependent
screening length $\Lambda^*<\Lambda$. Panel (b) presents the $%
\varphi(\rho)$ plots calculated for the realistic InO/SiO$_{2}$ structure
and different distances to the gate. One sees how the potential acquires
more and more local character as the gate approaches the film surface.

To investigate the $\varphi (\rho )$ dependence inside the film in
detail, we find the analytical solution to the system (\ref%
{Poisson}). For distances $\rho $ larger than the film thickness $d$ and for 
$\kappa \gg \kappa _{a},\kappa _{b}$ the potential is given by (see
Methods): 
\begin{equation}
\varphi (\rho )=2\frac{q}{\kappa d}\int\limits_{0}^{\infty }\frac{%
J_{0}\left( k\rho \right) }{k+\frac{\kappa _{a}\coth \left( kh\right)
+\kappa _{b}}{\kappa d}}dk.  \label{main_eq}
\end{equation}%
Here $J_{0}$ is the zero order Bessel function. Shown in Fig.\thinspace \ref%
{fig:potential}a is the semi-log plot of the potential vs. the distance
calculated for the same parameters as in Fig.\thinspace \ref{fig:model}b. We
clearly observe the change of behaviour from the logarithmic one
to the fast decay at longer distances. The corresponding screening length
at which the crossover occurs, $\Lambda^*$, is evaluated via the abscissa
section by the straight line corresponding to $\varphi (\rho )\propto \ln
(\rho /\Lambda^*)$ at small $\rho $. 
Plotting $\Lambda^*$ vs. $h$ in a
double-log scale (Fig.\thinspace \ref{fig:potential}b) we find $\Lambda^*\propto \sqrt{h}$ at $h\lesssim 10^{-1}\Lambda $. 
At larger $h$, the $\Lambda^*(h)$ dependence starts to deviate from the square root behaviour, and, eventually,
at sufficiently large $h$ the influence of the gate vanishes and $\Lambda^*$ saturates to $\Lambda$. 
Inspecting more carefully the transition region around $h\sim 10^{-1}\Lambda $, one observes that the functional dependence of the screened potential changes its character. 
At these scales the potential is pretty well described as $\varphi (\rho )\propto \exp
(-\rho /\Lambda^*)$ with the same $\Lambda^*\propto \sqrt{h}$ (see
Fig.\thinspace \ref{fig:potential}a) at $h\lesssim 10^{-1}\Lambda $. At $%
h\gtrsim 10^{-1}\Lambda $ the potential decays as a power $\varphi (\rho
)\propto \rho ^{-n}$, with $n\lesssim 3$.

To gain insight into the observed behaviours of the potential, we undertake the detailed analysis of two asymptotic cases, $\rho > h$ and $\rho < h$, in which the exact formulae for $\varphi(\rho)$ can be obtained.
Considering possible relations between $h$ and other relevant spatial scales, we derive, with the logarithmic accuracy, the  asymptotic behaviours of $\varphi(\rho)$ for corresponding sub-cases (see Methods for the details of calculations).
Our findings are summarized in Table\,\ref{tab:asymptotes}.

(A) At distances less than the film-electrode separation, $\rho < h$, we assume that $\coth \left( kh\right)\simeq 1$ in Eq.\,(\ref{main_eq}) and  recover the well-known result for the system without gate\,\cite{Rytova1967,chaplik1971charged,keldysh1979coulomb}:
\begin{equation}
\varphi (\rho )=\pi \frac{q}{\kappa d}\Phi _{0}\left( \frac{\rho }{\Lambda }%
\right),  \label{Rytova}
\end{equation}%
where $\Phi _{0}(x)=H_{0}\left( x\right) -N_{0}\left( x\right) $ is the
difference of the zero order Struve and Neumann functions \cite{Abramowitz}. Making use of the asymptotes for $\Phi_{0}$ given in Methods we find that at short distances $\rho<\Lambda$ one obtains logarithmic behavior of Eq.\,(\ref{Rytova}), while at large distances the field lines leave the film and one has the 3D Coulomb decay of the potential. 

(B) For $\rho > h$ we find 
\begin{equation}
\varphi (\rho )=\pi \frac{q}{\kappa d}\frac{1}{\xi _{1}-\xi _{2}}\left[ \xi
_{1}\Phi _{0}\left( \xi _{1}\frac{\rho }{\Lambda }\right) -\xi _{2}\Phi
_{0}\left( \xi _{2}\frac{\rho }{\Lambda }\right) \right] ,\quad \mathrm{where%
}\quad \xi _{1,2}=\frac{1}{2 (\kappa_a+\kappa_b) }\left[ \kappa _{b}\pm 
\sqrt{\kappa _{b}^{2}-4\kappa _{a}\kappa \,d/h}\right].  \label{smallh}
\end{equation}

Depending on  $h$, the length-scaling parameters, $\xi_1$ and $\xi_2$ can be either the real numbers, if  $h> 4d\kappa \kappa _{a}/\kappa _{b}^{2} $, or the complex mutually conjugated numbers, if $h< 4d\kappa \kappa _{a}/\kappa _{b}^{2} $.   This leads to the different regimes of the potential decay (see Table 1) that are controlled by the new screening lengths,  $\Lambda _{1,2}=\Lambda/\xi _{1,2}$ ($\Lambda _{1}<\Lambda _{2}$) in the former case and $\Lambda _{3}=\Lambda /\left\vert \xi_{1}\right\vert =\Lambda /\left\vert \xi_{2}\right\vert$ in the latter one. 
In particular, the logarithmic behaviour presented in sections (iii) and (vi) of Table~1, perfectly reproduces the results of computations shown in Fig.\thinspace \ref{fig:potential}a. 
For small $h< 4d\kappa \kappa _{a}/\kappa _{b}^{2} $ the empirical screening length $\Lambda^\ast$, acquires the form  $\Lambda _{3}=\sqrt{\left( \kappa /\kappa _{a}\right) dh}$ corresponding to the  small-$h$ square-root behaviour inferred from the curve of Fig.\,\ref{fig:potential}b. For $h> 4d\kappa \kappa _{a}/\kappa _{b}^{2} $ the logarithmic behaviour persists but with $\Lambda^\ast=\Lambda_1$, which saturates to $\Lambda$ with growing thickness of the spacer, $h$, between the film and the gate. 

At large scales above $\Lambda^\ast$, the screened charge potential decays following the power law, $\varphi(\rho)\propto\rho^{-n}$, where the exponent varies from $n=1$ (3D Coulomb charge interaction) to  $n=3$ (dipole-like interaction), in accord with the computational results discussed above. Which of the scenarios is realized, depends on the ratio of $\rho$ to $\Lambda_1$, $\Lambda_2$, and $\Lambda_3$, see Table\,1. Finally, for the small spacer thickness, the power-law screening transforms into the exponential one,  $\varphi(\rho)\propto\frac{2q}{\kappa d}\sqrt{\frac{\pi }{2}\frac{\Lambda _{3}}{\rho }}e^{-\rho /\Lambda _{3}}$, see Methods. This evolution  is well seen in the Fig.\,\ref{fig:potential}a, as improving fits of the potential curves to the exponential dependencies (shown by dashed lines) upon decreasing $h$.

{\renewcommand{\arraystretch}{1.7}
\begin{table}[h!]
\centering 
\begin{tabular}{|c|c|}
\hline
\multicolumn{2}{|c|}{$\rho < h$} \\ \hline
& 
\begin{tabular}{c|c}
\begin{tabular}{c}
(i) \,\,\,\,\,\,\,
$\rho <\Lambda $ \\ 
$\varphi (\rho )\simeq -2\frac{q}{\kappa d}\ln \frac{C\rho }{2\Lambda }$
\vspace{3mm}%
\end{tabular}
& 
\begin{tabular}{c}
(ii)  \,\,\,\,\,\,\,  
$\rho >\Lambda $ \\ 
$\varphi (\rho )\simeq 2\frac{q}{(\kappa_a+\kappa_b)\rho }$
\vspace{3mm}
\end{tabular}%
\end{tabular}
\\ \hline\hline
\multicolumn{2}{|c|}{$\rho > h$} \\ \hline
$h>4d\kappa \kappa _{a}/\kappa _{b}^{2}$ & 
\begin{tabular}{c|c|c}
\begin{tabular}{c}
(iii) \,\, 
$\rho <\Lambda _{1}<\Lambda _{2}$ \\ 
$\varphi \simeq -2\frac{q}{\kappa d}\ln \frac{C\rho }{2\Lambda_1}$
\vspace{3mm}%
\end{tabular}
& 
\begin{tabular}{c}
(iv) \,\, 
$\Lambda _{1}<\rho <\Lambda _{2}$ \\ 
$\varphi \simeq \frac{2}{(\kappa _{b}^{2}-4\kappa _{a}\kappa \,d/h)^{1/2}}%
\frac{q}{\rho }$
\vspace{3mm}%
\end{tabular}
& 
\begin{tabular}{c}
(v) \,\,
$\Lambda _{1}<\Lambda _{2}<\rho $ \\ 
$\varphi \simeq 2 \frac{\kappa _{b}}{\kappa _{a}^{2}}\frac{qh^{2}}{\rho ^{3}}$
\vspace{3mm}%
\end{tabular}%
\end{tabular}
\\ \hline
$h< 4d\kappa \kappa _{a}/\kappa _{b}^{2}$  & 
\begin{tabular}{c|c}
\begin{tabular}{c}
(vi) \,\, 
$\rho <\Lambda _{3}$ \\ 
$\varphi \simeq -2\frac{q}{\kappa d}\ln \frac{C\rho }{2\Lambda_3}$
\vspace{3mm}%
\end{tabular}
& 
\begin{tabular}{c}
(vii) \,\, 
 $\rho >\Lambda _{3}$ \\ 
$\varphi \simeq 2\frac{\kappa _{b}}{\kappa _{a}^{2}}\frac{qh^{2}}{\rho ^{3}}$
\vspace{3mm}%
\end{tabular}%
\end{tabular}
\\ \hline
\end{tabular}
\caption{\textbf{Regimes of the interaction.}
There are two major regions, short distances,  $\rho<h$, where interaction is only weakly influenced by the gate (upper panel),  and large distances, $\rho>h$, where the gate presence renormalizes the interaction (bottom panel). Logarithmic dependence on $\rho$ appears  below the respective screening lengths, $\Lambda$, $\Lambda_1$ and $\Lambda_3$. Above these lengths the potential decays according to the power law. The constant 
$C=e^{\gamma }\simeq 1.781...$ is the exponent of the Euler constant $\gamma$.}
\label{tab:asymptotes}
\end{table}
}

\section*{Discussion}

The above results describe a wealth of electrostatic regimes in which the high-$\kappa$ sheets can operate depending on the distance to the control gate.
The interrelation between the regimes presented in the Table 1 is conveniently illustrated in Fig.\,\ref{fig:potential}c
showing the map of the interaction regimes drawn for the InO/SiO$_2$ heterostructure parameters. Note that the specific structure of the map depends on the particular values of the parameters of the system controlling the ratios between the different screening lengths $\Lambda$, $\Lambda_1$, $\Lambda_2$, and $\Lambda_3$. The lines visualizing these lengths mark crossovers between different interaction regimes. The gray roman numerals correspond to the regimes listed in the Table\,1. 
The colors highlight the basic functional forms of interactions between the charges.  
The bluish area marks the manifestly high-$\kappa$ regions of the unscreened 2D logarithmic Coulomb interaction. As the distance to the gate becomes less than the
separation between the interacting charges, the screening length restricting the logarithmic interaction regimes
renormalizes from $\Lambda$ to either $\Lambda_1$ or $\Lambda_3$.
The line $\Lambda_2$ delimits the large-scale  point-like and dipolar-like interaction regimes. 
At very small $h$, a petal-shaped region appears in which the potential drops exponentially with the distance at $\rho>\Lambda_3$.

The implications of the tunability of the logarithmic Coulomb interactions are far reaching. The charge logarithmic confinement is the foundation of the charge BKT transition. 
Thus tuning the range of the confinement offers a perfect laboratory for the study of effects of screening on the BKT transition and related phenomena. Most notably, adjusting the gate spacer, one can can regulate the effects of diverging dielectric constant near the metal- and superconductor-insulator transitions\,\cite{BatVin2013}.  Addressing the technological applications, we envision a wide use of gate controlled electrostatic screening in the high-$\kappa$ films-based flash memory circuits.
The reduction of the Coulomb repulsion from the 2D long-range logarithmic to the point- or dipolar- and even to the exponential ones will crucially scale down the circuit size, increasing their capacity and reliability.


\section*{Methods}

\subsection*{Fourier transformation}

We seek the solution of equations (\ref{Poisson}) in the form: 
\begin{gather}
\varphi _{a}=\int\limits_{0}^{\infty }A_{1}\left( k\right)
e^{-kz}J_{0}\left( k\rho \right) dk+\int\limits_{0}^{\infty }A_{2}\left(
k\right) e^{kz}J_{0}\left( k\rho \right) dk;  \label{System} \\
\varphi =\frac{q}{\kappa }\int\limits_{0}^{\infty }e^{-k\left\vert
z\right\vert }J_{0}\left( k\rho \right) dk+\int\limits_{0}^{\infty
}B_{1}\left( k\right) e^{-kz}J_{0}\left( k\rho \right)
dk+\int\limits_{0}^{\infty }B_{2}\left( k\right) e^{kz}J_{0}\left( k\rho
\right) dk;  \notag \\
\varphi _{b}=\int\limits_{0}^{\infty }D\left( k\right) e^{kz}J_{0}\left(
k\rho \right) dk.  \notag
\end{gather}

Making use the specified in the text electrostatic boundary conditions we
get a system of linear equations for coefficients $A_{1,2}$, $B_{1,2}$ and $%
D $: 
\begin{gather}
\frac{q}{\kappa }+B_{1}+B_{2}e^{kd}=A_{1}+A_{2}e^{kd},\qquad \frac{q}{\kappa 
}+B_{1}-B_{2}e^{kd}=\frac{\kappa _{a}}{\kappa }A_{1}-\frac{\kappa _{a}}{%
\kappa }A_{2}e^{kd},  \label{Lin} \\
\frac{q}{\kappa }+B_{1}e^{kd}+B_{2}=D,\qquad \frac{q}{\kappa }%
-B_{1}e^{kd}+B_{2}=\frac{\kappa _{b}}{\kappa }D,\qquad
A_{1}+A_{2}e^{2kh}e^{kd}=0.  \notag
\end{gather}

In particularly, for $B_{1,2}$ we obtain: 
\begin{equation}
B_{1,2}=-\frac{q}{\kappa }\frac{\beta _{1,2}\left( \beta
_{2,1}+e^{kd}\right) }{\beta _{1}\beta _{2}-e^{2kd}},  \label{Coeff}
\end{equation}%
with 
\begin{equation}
\beta _{1}=\frac{1-{\kappa _{b}}/{\kappa }}{1+{\kappa _{b}}/{\kappa }}\quad 
\mathrm{and}\quad \beta _{2}=\frac{\tanh kh-{\kappa _{a}}/{\kappa }}{\tanh
kh+{\kappa _{a}}/{\kappa }}.  \label{betas}
\end{equation}

We are interested in distances, $\rho$, larger than the film thickness $d$
when the main contribution to integrals (\ref{System}) is coming from $k\ll
d^{-1}$. \ Expanding (\ref{Coeff}) over the small parameter $kd$, assuming
that $\kappa \gg \kappa _{a},\kappa _{b}$ in (\ref{betas}) and substituting
the resulting coefficients $B_{1,2}$ into the integral for $\varphi $ in (%
\ref{System}) we obtain the expression (\ref{main_eq}).

\subsection*{Integrals}

Integral (\ref{main_eq}) can be evaluated using the standard table integral 
\cite{gradshteyn2014table} 
\begin{equation}
\int\limits_{0}^{\infty }\frac{J_{0}\left( ak\right) }{k+z}dk=\frac{\pi }{2}%
\Phi _{0}\left( az\right)  \label{Integral}
\end{equation}%
(here $z=x+iy$  is the complex variable) in two limit cases.

i) In the limit $\rho < h$ the main contribution to (\ref{main_eq}) comes
from the high-$k$ values and $kh\gg 1$. Assuming $\coth \left( kh\right)
\simeq 1$ we reduce (\ref{main_eq}) to (\ref{Integral}) and obtain the
expression (\ref{Rytova}).

ii) In the limit $\rho > h$ the main role is played by the low-$k$ region, 
$kh\ll 1$. Then $\coth \left( kh\right) \simeq 1/kh$ and the integral (\ref{main_eq}) can be calculated by partial fraction decomposition onto two
integrals,

\begin{equation}
\varphi (\rho )\simeq 2\frac{q}{{\kappa }d}\frac{1}{\xi _{1}-\xi _{2}}\left[
\int\limits_{0}^{\infty }\frac{\xi _{1}J_{0}\left( k\rho \right) }{k+\xi
_{1}\Lambda ^{-1}}dk-\int\limits_{0}^{\infty }\frac{\xi _{2}J_{0}\left(
k\rho \right) }{k+\xi _{2}\Lambda ^{-1}}dk\right], \label{calc2Int}
\end{equation}%
where $\xi _{1}$ and $\xi _{2}$ are the given by (\ref{smallh}) solutions of
the characteristic quadratic equation $\Lambda \xi ^{2}+\gamma _{b}\xi
+\gamma _{a}h^{-1}=0$. Each of these integrals is of the type (\ref{Integral}%
) that permit us to obtain~(\ref{smallh}).

\subsection*{Limit expansions}

The asymptotic expansions of $\Phi _{0}$ as a function of the complex argument, $z=x+iy$, are found from the table properties of $H_{0}$ and $N_{0}$ \, \cite{Abramowitz}. When $z\rightarrow 0$ the function $\Phi _{0}$  can be approximated as $\Phi _{0}\left( z\right) \simeq -\frac{2}{\pi }\ln \frac{1}{2}Cz$ where $C=e^{\gamma }\simeq 1.781$ is the exponent of the Euler constant. At large $\left\vert z\right\vert \gg 1$ the Laurent series development $\Phi _{0}\left( z\right) \simeq \frac{2}{\pi }\left( z^{-1}-z^{-3}\right) $ is suitable over the whole complex plane except the vicinity of the imaginary axis $z=iy$, where the real part of this expansion vanishes and the non-analytic contribution prevails. The latter can be accounted for, by presenting $\mathrm{Re}\,\Phi _{0}\left( iy\right) $ via the Macdonald function $K_{0}$, $\mathrm{Re}\,\Phi _{0}\left( iy\right) =\frac{2}{\pi }K_{0}\left( y\right) $ that is approximated as $\simeq \sqrt{\frac{2}{\pi y}}e^{-y}$ at $y\gg 1$.


\noindent

\section*{Acknowledgements }

This work was supported by ITN-NOTEDEV FP7
mobility program. The work by V.V. and partly I.L. was supported by the U.S. Department of Energy, Office of Science,
Materials Sciences and Engineering Division. The work of T.I.B. was supported by the Ministry of Education and Science of the Russian Federation and by RSCF (project No 14-22-00143).
T.I.B. acknowledge for financial support the Alexander von Humboldt Foundation.

\section*{Author contributions statement}

T.I.B., I.L., and V.V. conceived the work. S.K., I.L., and V.V. performed calculations. T.I.B. contributed to analysing and presenting the data. 
All authors contributed to writing the manuscript.

\section*{Additional information}

\textbf{Competing financial interests} The authors declare that they have no competing financial interests.


\end{document}